\documentclass[aps,showpacs,showkeys,prl,twocolumn,superscriptaddress,floatfix]{revtex4}
\usepackage{pstricks,pst-node,pst-text,pst-3d,pst-plot,amsfonts,amssymb}
\usepackage{graphicx}
\usepackage{dcolumn}
\usepackage{subfigure}

\usepackage{color}

\newcommand{\Ts}{T_{\textrm{\scriptsize s}}}

\newcommand{\Tc}{T_{\textrm{\scriptsize c}}}
\newcommand{\Qc}{Q_{\textrm{\scriptsize c}}}
\newcommand{\Mc}{M_{\textrm{\scriptsize c}}}
\newcommand{\deltac}{\Delta_{\textrm{\scriptsize c}}}
\newcommand{\chic}{\chi_{\textrm{\scriptsize c}}}
\newcommand{\SO}{\textrm{SO(2)}}
\newcommand{\Z}{\mathbb{Z}_2}

\definecolor{blue}{rgb}{0,0,1}
\definecolor{red}{rgb}{1,0,0}
\definecolor{green}{rgb}{0,1,0}

\def\latticeFig {\includegraphics[width=8cm]{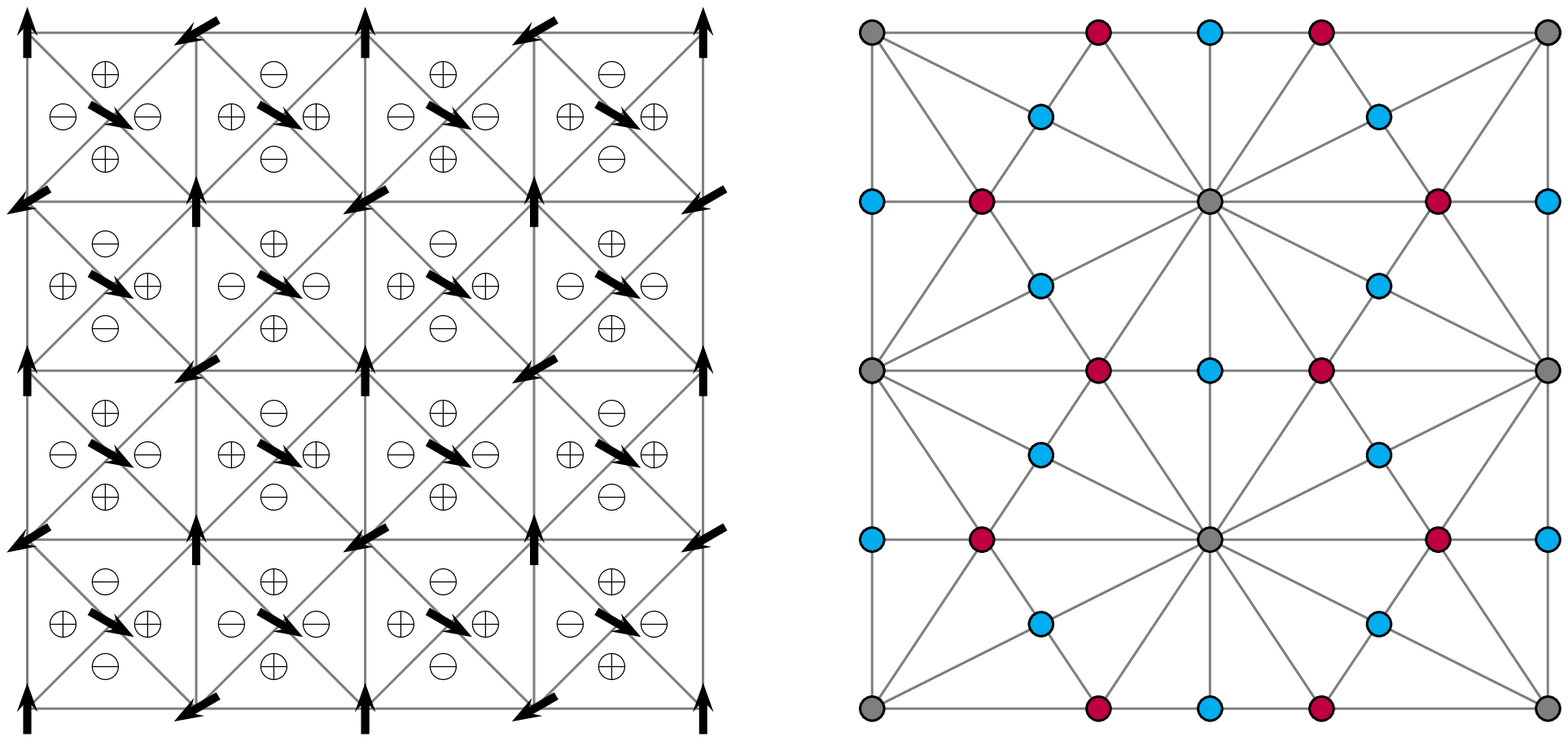}}
\def\summaryFig {\includegraphics[trim=27 642 340 70, clip,width=8cm]{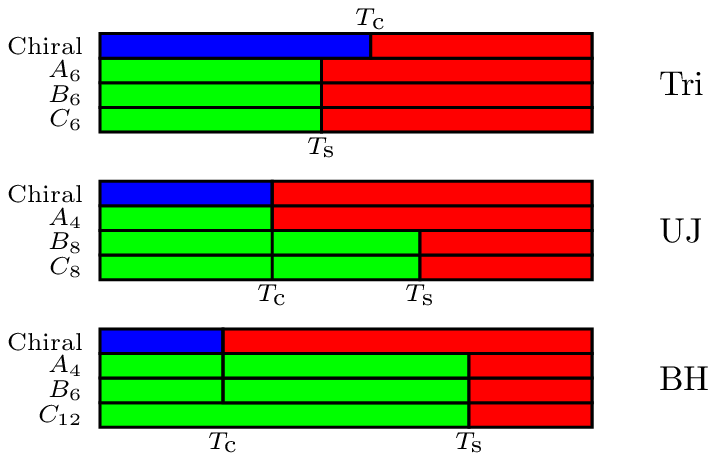}}
\def\UJQBfig {\includegraphics[width=8cm]{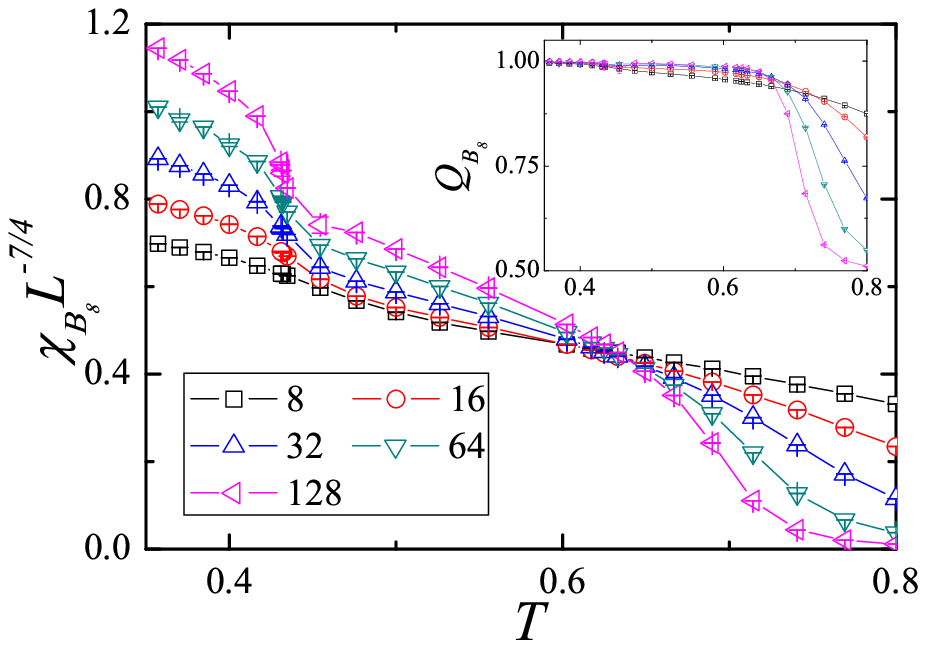}}
\def\UJQcFig  {\includegraphics[width=8cm]{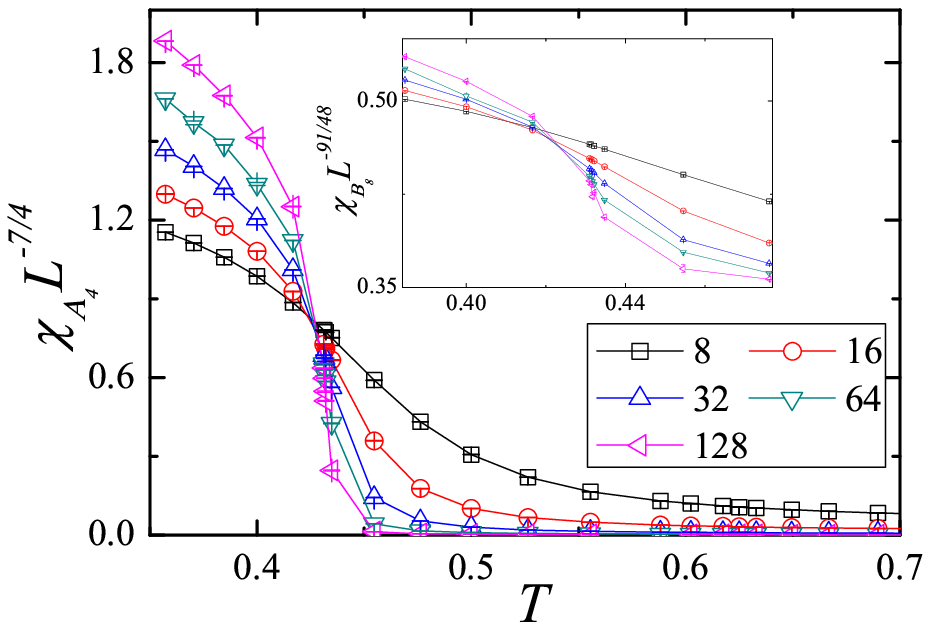}}
\def\UJCBfig{\includegraphics[width=8cm]{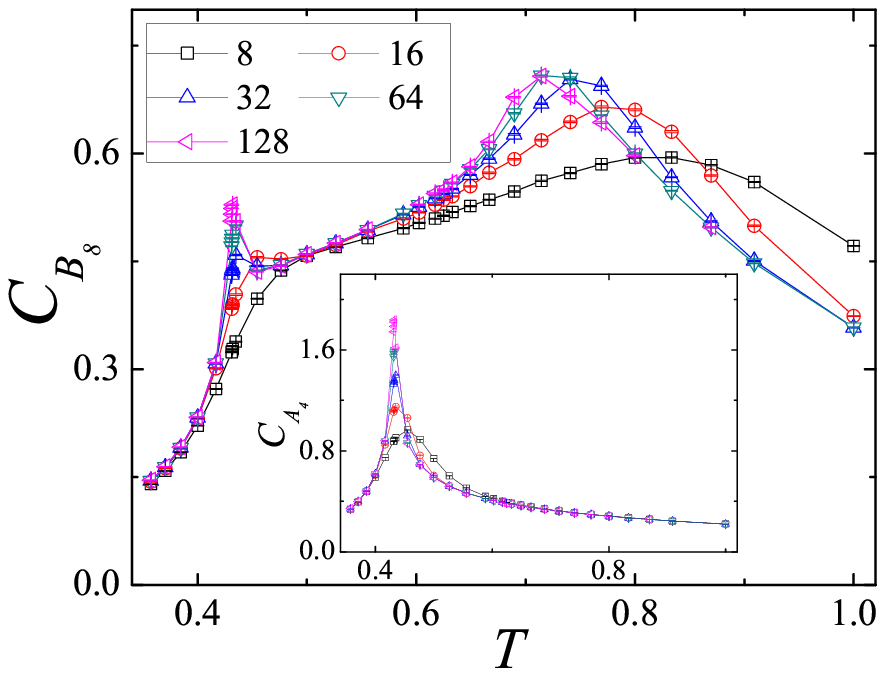}}
\def\UJchiBfig {\includegraphics[width=8cm]{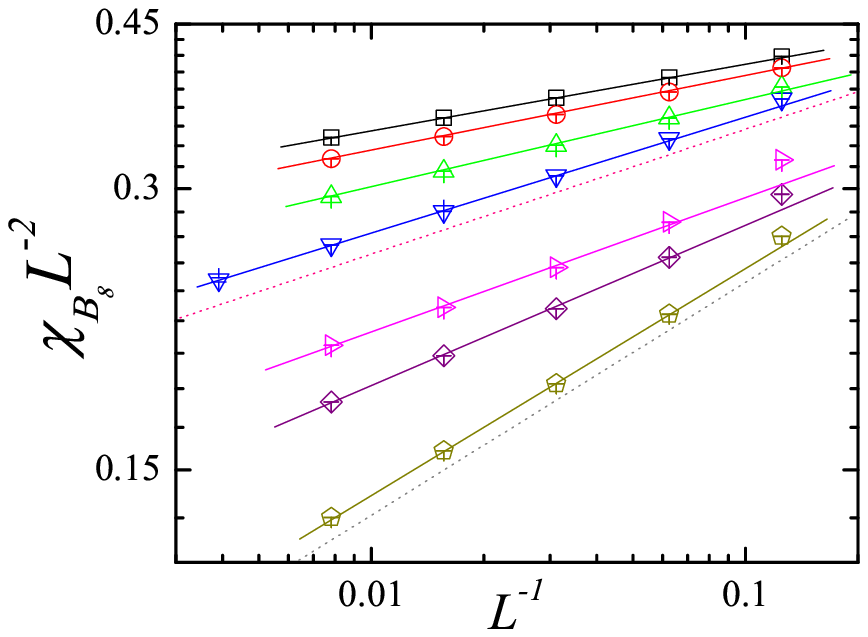}}
\def\BHCfig {\includegraphics[width=8cm]{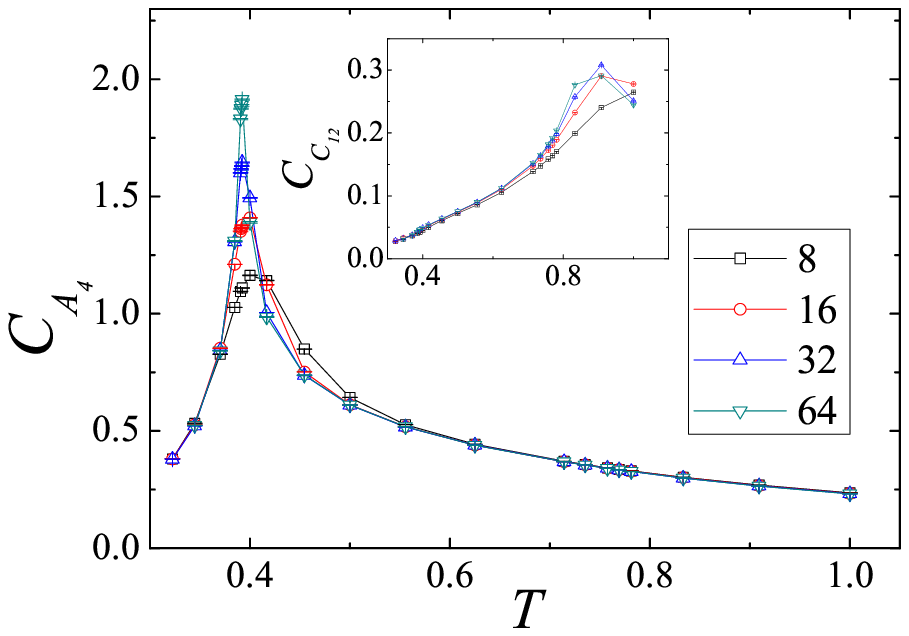}}

\begin{document}

\title{Novel phase transitions in XY Antiferromagnets on Plane Triangulations}
\author{Jian-Ping Lv}
\affiliation{Department of Physics, China University of Mining and Technology, Xuzhou 221116, China}
\author{Timothy M. Garoni}
\email{tim.garoni@monash.edu}
\affiliation{School of Mathematical Sciences, Monash University, Clayton, Victoria~3800, Australia}
\author{Youjin Deng}
\email{yjdeng@ustc.edu.cn}
\affiliation{Hefei National Laboratory for Physical Sciences at Microscale and Department of Modern Physics, University of Science and Technology of China, Hefei, Anhui 230026, China}

\begin{abstract}
Using Monte Carlo simulations and finite-size scaling, we investigate the XY antiferromagnet on the triangular, Union Jack and bisected-hexagonal lattices,
and in each case find both Ising and Kosterlitz-Thouless transitions.
As is well-known, on the triangular lattice, as the temperature decreases the system develops chiral order for temperatures $T < \Tc$,
and then quasi-long-range magnetic order on its sublattices when $T < \Ts$, with $\Ts < \Tc$.
The behavior $\Ts<\Tc$ is predicted by theoretical arguments due to Korshunov~\cite{Korshunov02}, based on the unbinding of kink-antikink pairs.
On the Union Jack and bisected-hexagonal lattices, by contrast, we find that as $T$ decreases the magnetizations on some of the sublattices become quasi-long-range ordered at a temperature $\Ts > \Tc$,
before chiral order develops.
In some cases, the sublattice spins then undergo a second transition, of Ising type, separating two quasi-long-range ordered phases.
On the Union Jack lattice, the magnetization on the degree-4 sublattice remains disordered until $\Tc$ and then undergoes an Ising transition to a quasi-long-range ordered phase.
\end{abstract}



\maketitle
Two-dimensional fully-frustrated XY (FFXY) models have been the subject of considerable interest over the past three decades; for a recent review see~\cite{HasenbuschPelissettoVicari05JSTAT}.
In addition to their intrinsic theoretical importance within the field of critical phenomena, such models can also be realized experimentally via Josephson Junction Arrays
(JJAs) in a uniform magnetic field~\cite{LingLezecHigginsTsaiFujitaNumataNakamuraOchiaiTangChaikinBhattacharya96,MartinoliLeemann00}.

On the triangular lattice, or indeed any triangulation of the plane, the FFXY model coincides with the usual antiferromagnetic XY (AFXY) model~\cite{LeeJoannopoulosNegeleLandau84,MiyashitaShiba84}, defined by
the reduced Hamiltonian
\begin{equation}
  \mathcal{H} = J\sum_{ij}\mathbf{s}_i \cdot \mathbf{s}_j = J\sum_{ij} \cos(\theta_i-\theta_j),
  \label{hamiltonian}
\end{equation}
where $\mathbf{s}_i=(\cos \theta_i, \sin \theta_i)$ is a planar spin with unit length on lattice site $i$, $J=1/T$ is the inverse temperature, and the summation is over all nearest-neighbor pairs of sites.

\begin{figure}
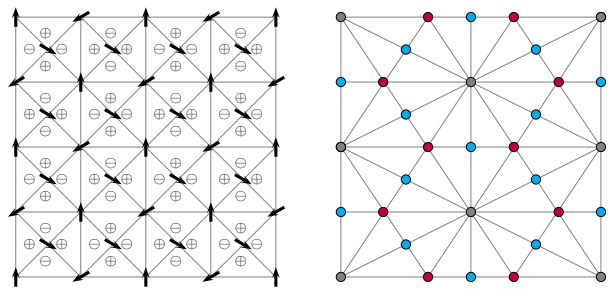

  \begin{center}
  \latticeFig
  \caption{\label{UJ} (Color online)
    (Left) A ground-state configuration of the AFXY model on the UJ lattice.
    Symbols `$+$' and `$-$' denote the chirality of the elementary triangular faces.
    (Right) A BH lattice. The tripartition of the vertex set is shown in purple/cyan/gray.}
  \end{center}
\end{figure}

The ground states of the triangular-lattice AFXY model are such that the spins on each of the three equivalent triangular sublattices are perfectly aligned,
with adjacent spins differing by an angle $\pm 2\pi/3$.
In addition to an $\SO$ rotational degeneracy, generic of XY models, there is a discrete $\Z\cong\textrm{O(2)}/\textrm{SO(2)}$ reflection degeneracy, induced by frustration,
corresponding to the two possible {\em chiralities} of each elementary triangular face.
The chirality of a face refers to the sign of the rotation of the spins as one traverses the face counterclockwise; see Fig.~\ref{UJ}.
At positive temperatures, the Mermin-Wagner theorem~\cite{Mermin67} forbids the sublattice spins from ordering.
However, for low temperatures, the sublattice magnetizations exhibit quasi-long-range (QLR) order (algebraically decaying correlations), while the chiral degrees of freedom exhibit a genuine long-range order.

Despite some early
controversy~\cite{LeeJoannopoulosNegeleLandau84,MiyashitaShiba84,YosefinDomany85,GuptaDeLappBartrouniFoxBaillieApostolakis88,GranatoKosterlitzLee91,LeeKosterlitzGranato91,RamirezSantiagoJose92},
there is now a consensus~\cite{LeeLee94,Olsen95,XuSouthern96,LeeLee98,CapriottiVaiaCuccoliTognetti98,Korshunov02,HasenbuschPelissettoVicari05JSTAT,OkumuraYoshinoKawamura11,ObuchiKawamura12}
that for a number of FFXY models sharing the same ground-state degeneracies, including the triangular-lattice AFXY model and square-lattice FFXY model~\cite{Villain77a,Villain77b,TeitelJayaprakash83},
the phase transition associated with the magnetic order parameter occurs at a temperature $\Ts$ strictly below the transition point $\Tc$ for the chiral order parameter.
It is generally accepted~\cite{HasenbuschPelissettoVicari05JSTAT} that the sublattice magnetizations disorder via a standard Kosterlitz-Thouless (KT)
transition~\cite{Berezinskii71,KosterlitzThouless73,Kosterlitz74}, while the chiralities undergo an Ising transition.
The behavior $\Ts<\Tc$ is supported by theoretical arguments based on the unbinding of kink-antikink pairs~\cite{Korshunov02,OlssonTeitel05}.

In this Letter, we study the AFXY model on the Union Jack (UJ) and bisected-hexagonal (BH) lattices, using Monte Carlo simulations and finite-size scaling.
The UJ and BH lattices are plane triangulations that share many of the properties of the triangular lattice; see Fig.~\ref{UJ}.
In particular, the ground-state degeneracies of the AFXY model on each of these three lattices are identical.
However, as we show, the critical behavior of the AFXY model on the UJ and BH lattices is qualitatively different to the triangular-lattice case.
In particular, we find that on both the UJ and BH lattices $\Ts >\Tc$.

Like the triangular lattice, the UJ and BH lattices are tripartite; each consists of three independent sublattices, which we label $A$, $B$, $C$.
Since each sublattice is regular, we use a subscript to indicate its coordination number.
The triangular lattice consists of sublattices $A_6$, $B_6$ and $C_6$, UJ consists of $A_4$, $B_8$, $C_8$, and BH consists of $A_4$, $B_6$, $C_{12}$; see Fig.~\ref{UJ}.
Unlike the triangular lattice, the sublattices of the UJ and BH lattices are not all equivalent, and this leads to interesting new physics.

On the UJ lattice, we find that the spins on the $A_4$ sublattice become QLR-ordered at a different temperature to the spins on the $B_8$ and $C_8$ sublattices.
Analogous behavior has very recently been observed in the 4-state Potts antiferromagnet~\cite{ChenQinChenWeiZhaoNormandXiang11,DengHuangJacobsenSalasSokal11}.
Perhaps more surprisingly, the magnetic transition of the $A_4$ spins, separating the disordered and QLR-ordered phases, appears to be an Ising transition.
Such novel Ising transitions leading to QLR order, rather than genuine long-range order, have very recently been observed~\cite{ShiLamacraftFendley11} in generalized XY models
whose Hamiltonians contain nematic-like interactions~\cite{LeeGrinstein85,ParkOnodaNagaosaHan08,PoderosoArenzonLevin11,ShiLamacraftFendley11}.
To our knowledge, our study presents the first observation of an Ising magnetic transition in the standard XY antiferromagnet.

\begin{figure}
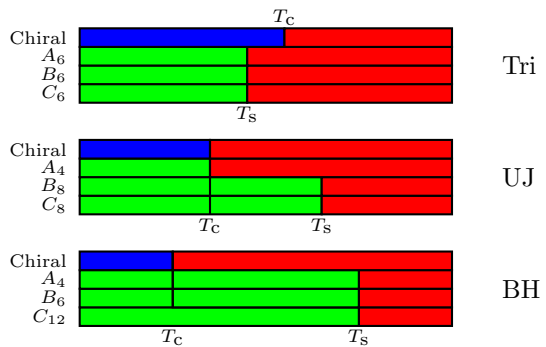

  \begin{center}
    \summaryFig
    \caption{\label{summary} (Color online) Summary of phases and transitions for AFXY model on triangular (Tri), Union Jack (UJ) and bisected-hexagonal (BH) lattices.
      Red refers to disorder, blue to order, and green to QLR order.
      Temperature increases from left to right.}
  \end{center}
\end{figure}

\paragraph{Summary of Results.---}
A summary of the qualitative behaviour of each lattice is presented in Fig.~\ref{summary}.
On each lattice, we observe a chiral transition at a temperature $T=\Tc$, with $\Tc(\textrm{Tri})>\Tc(\textrm{UJ})>\Tc(\textrm{BH})$.
In each case, we find strong evidence that the chiral transition is in the Ising universality class.
In addition, in each case we also observe magnetic spin transitions at a temperature $\Ts\neq \Tc$, with $\Ts(\textrm{Tri})<\Ts(\textrm{UJ})<\Ts(\textrm{BH})$.
The qualitative features of the magnetic transition are highly lattice-dependent however.
On the triangular lattice, we observe a magnetic transition from disorder to QLR order on each of the three equivalent sublattices at $\Ts>\Tc$.
The transition is consistent with the KT universality class.

On the UJ lattice, we observe that the spins on sublattices $B_8$ and $C_8$ undergo a transition from disorder to QLR order, which is again consistent with KT behaviour.
However on the UJ lattice, we find very clear evidence that $\Ts>\Tc$.
Furthermore, we find no evidence of a spin transition on the $A_4$ sublattice at $\Ts$,
but we find strong evidence that the spins on the $A_4$ sublattice undergo an Ising transition from disorder to QLR order at (or extremely close to) $\Tc$.
In addition, it appears that the spins on the $B_8$ and $C_8$ sublattices undergo a second transition at $\Tc$, which separates two QLR-ordered phases which have different exponents but are otherwise equivalent.
This second transition appears to be in the Ising universality class, but the measured magnetic exponent takes a non-Ising value of $1.94(1)$.

On the BH lattice, the spins on each sublattice $A_4$, $B_6$ and $C_{12}$, appear to transition from disorder to QLR order at a common value of $\Ts>\Tc$.
The transition appears to be consistent with the KT universality class.
The spins on $A_4$ and $B_6$ then undergo a second transition, consistent with the Ising universality class, at $T=\Tc$, while the spins on $C_{12}$ do not.
The $C_{12}$ sublattice appears to display generic KT behaviour.

\paragraph{Monte Carlo Simulations.---}
Our simulations used a local algorithm based on a mixture of standard local Metropolis~\cite{LandauBinder09} updates together with overrelaxed~\cite{BrownWoch87,Creutz87} (microcanonical) updates.
Periodic boundary conditions are applied, and the maximum linear system size is $L_{\textrm{\scriptsize max}}=384,256$ and $256$ for the triangular, UJ, and BH lattices, respectively.
The total CPU time for the simulations was approximately $60$ years, with a $3.2$ GHz Xeon EM64T processor.

For each sublattice $S=A,B,C$, we define the magnetic order parameter $\mathbf{M}_S = (1/V_S) \sum_{i \in S} \mathbf{s}_i$, with $V_S=|S|$.
The chiral order parameter is defined as $$\Mc=\frac{2}{3\sqrt{3}N_\bigtriangleup} \sum_{\bigtriangleup}[\sin(\theta_{i}-\theta_{j})+\sin(\theta_{j}-\theta_{k})+\sin(\theta_{k}-\theta_{i})],$$
where the summation is over the $N_{\bigtriangleup}$ triangular faces.
The sequence $(i,j,k)$ is alternately chosen clockwise and counterclockwise on neighboring triangles.
The magnetic and chiral susceptibilities are then defined to be $\chi_{S}= V_S \langle M_S^2\rangle$ and $\chic=N_\bigtriangleup \langle \Mc^2\rangle$,
and we also define dimensionless ratios $Q_{S}= \langle M_{S}^2 \rangle^2 / \langle M_{S}^4\rangle$ and $\Qc=\langle \Mc^2\rangle^2/\langle \Mc^4\rangle$.
In addition, on each sublattice $S$, we measure the nearest-neighbor spin-spin correlation function $E_S$, and define a specific-heat-like quantity $C_{S}=V_S(\langle E_S^2 \rangle-\langle E_S \rangle^2)$.

\paragraph{Triangular lattice.---}
As $T$ decreases, the $Q_c$ data for different system sizes display an approximately common intersection
near $T \approx 0.512$.
For $T\lessapprox0.512$, the $\Qc$ value quickly approaches $1$, and $\langle \Mc^2 \rangle $
converges to a nonzero value as $L$ increases, implying the occurrence of chiral order.
The spins on the three sublattices remain disordered until $T\approx 0.50$,
where the $Q_{S}$ data indicate a magnetic transition on each of the three equivalent sublattices $S=A_6,B_6,C_6$.
In the low-temperature region $T\lessapprox0.50$, as $L$ increases, the $Q_{S}$ data converge to
a line of non-trivial $T$-dependent values, implying that each sublattice is QLR ordered.

For small $\deltac=T-\Tc$, the ratio $\Qc$ is expected to behave like
\begin{equation}
\Qc= \Qc^*+a_1 \deltac L^{y_t}+ a_2 \deltac^2 L^{2y_t} + b/L,
\label{Q ansatz}
\end{equation}
where $a_1$, $a_2$, and $b$ are free parameters, and $y_t = 1/\nu$ is the leading thermal renormalization exponent.
We performed least-squares fits of~(\ref{Q ansatz}) to the $Q_c$ data near $T\approx0.512$.
The data in the range $24 \leq L \leq 192$ are well-described by~(\ref{Q ansatz}), and the fit yields $\Tc = 0.5123(2)$, $y_t=0.99(1)$, and $\Qc^* =0.8587(3)$.

We also performed least-squares fits of the $\chic$ data to
\begin{equation}
  \chic = L^{2y_h-2} \left(a_0+a_1 \deltac L ^{y_t}+ a_2 \deltac^2 L^{2y_t} + b/L \right),
  \label{chi ansatz}
\end{equation}
with the fixed Ising value $y_t=1$.
This yields a critical point $\Tc=0.5123(1)$ and magnetic exponent $y_h=1.874(3)$.

The estimated critical exponents, $y_t=0.99(1)$ and $y_h=1.874(3)$, agree well with the exact results $y_t=1$ and $y_h=15/8$ for the two-dimensional Ising model.
The critical value $\Qc^* =0.8587(3)$ is also consistent with the existing result $Q_{\textrm{\scriptsize Ising}}^*=0.85872528(3)$ for the ferromagnetic triangular-lattice Ising model~\cite{KamieniarzBlote93}.
We conclude that the chiral transition is in the Ising universality class.
We note that our estimate of $y_t$ is inconsistent with several earlier values reported in the literature~\cite{LeeLee98,OzekiIto03,ObuchiKawamura12} which suggested $y_t\approx1.2$.
Our estimate of $\Tc$ is close to the recent estimate~\cite{ObuchiKawamura12} of $\Tc=0.51254(3)$.

Compared with the chiral transition, it is much more difficult to locate the spin transition of the sublattice magnetizations.
Fitting the ansatz (\ref{Q ansatz}) to the $Q_{A_6}$ data we obtain $\Ts=0.50(1)$.
It is known~\cite{KennaIrving97,Janke97} that at a KT transition the susceptibility diverges as $\chi\propto L^{7/4}(\ln L)^{1/8}$.
Assuming that the spin transition is of KT type, we therefore consider the curves of $\chi_{A_6} L^{-7/4} (\ln L)^{-1/8}$ versus $T$ for a number of fixed $L$.
From the scaling of the intersections for various $32\leq L\leq256$ and $0.500 \le T \le 0.505$, we obtain $\Ts=0.5040(3)$.
This value is consistent with the recent estimate~\cite{ObuchiKawamura12} of $\Ts=0.504(1)$.

\paragraph{Union-Jack lattice.---}
\begin{figure}
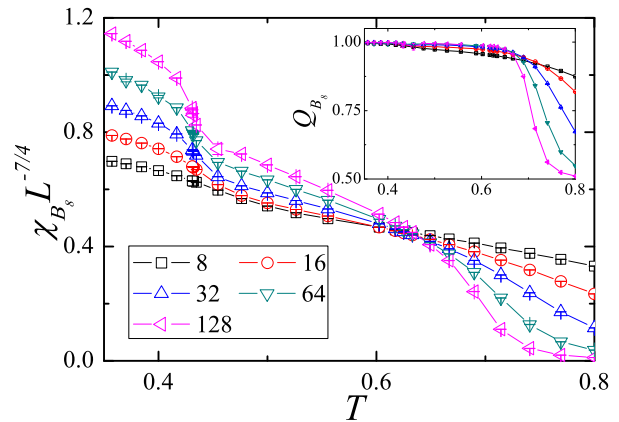

  \begin{center}
    \UJQBfig
    \caption{\label{UJ B8} (Color online) Quantity $\chi_{B_8} L^{-7/4}$ vs $T$ for various $L$, on the UJ lattice. The inset shows $Q_{B_8}$ vs $T$.}
  \end{center}
\end{figure}
\begin{figure}
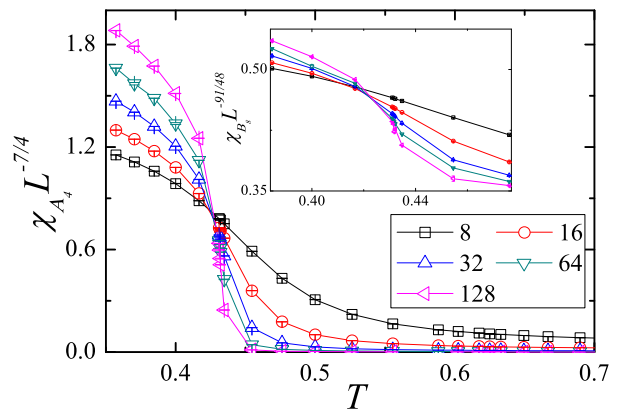

  \begin{center}
    \UJQcFig
    \caption{\label{UJ A4 and chiral} (Color online) Quantity $\chi_{A_4} L^{-7/4}$ vs $T$ for various $L$, on the UJ lattice. The inset shows $\chi_{B_8} L^{-91/48}$ vs $T$.}
  \end{center}
\end{figure}
From Fig.~\ref{UJ B8}, which plots $\chi_{B_8}L^{-7/4}$ and $Q_{B_8}$ vs $T$, we see that the spins on the sublattice $B_8$ (and also $C_8$) undergo two distinct phase transitions;
one at $T \approx 0.43$ and another at $T\approx0.64$.
By contrast, Fig.~\ref{UJ A4 and chiral} shows that the spins on sublattice $A_4$ undergo a single transition near $T \approx 0.43$.

Assuming that the spin transition near $T\approx0.64$ is a KT transition, the $\chi_{B_8}$ data were analyzed in an analogous manner to that described for the $\chi_{A_6}$ data on the triangular lattice.
This yields an estimate of the spin transition point on the $B_8$ and $C_8$ lattices of $\Ts=0.639(2)$.

For the chiral transition, fitting the $\Qc$ data to~(\ref{Q ansatz}) analogously to the triangular case yields $\Tc =0.4316(1)$, $y_t=0.99(1)$, and $\Qc^*=0.8562(3)$.
Similarly, fitting $\chic$ to~(\ref{chi ansatz}) with $y_t=1$ fixed yields $\Tc =0.4316(1) $ and $y_h=1.874(3)$.
This again suggests that the chiral transition is in the Ising universality class.

Perhaps surprisingly, fitting the $Q_{A_4}$ and $\chi_{A_4}$ data near $T\approx0.64$ in a similar way yields $T_{A_4} =0.4316(1)$, $y_t=1.00(1)$ and $y_h=1.874(2)$.
This suggests two things.
Firstly, within the resolution of our simulations, the transition points of the $A_4$ and chiral transitions coincide, implying that either $T_{A_4}=\Tc$ exactly, or that $|T_{A_4}-\Tc|\lessapprox 0.0001$.
Secondly, the critical behaviour associated with the $A_4$ spin transition is consistent with the Ising universality class.
This suggests that the magnetization on the $A_4$ sublattice undergoes an Ising transition from a disordered phase to a QLR-ordered phase at (or very close to) $T=\Tc$.
Based on these observations, we conjecture that the $A_4$ sublattice spins undergo an Ising transition precisely at the chiral transition temperature $\Tc$.

In addition, we find that the spins on sublattices $B_8$ and $C_8$ also undergo a transition near (or at) $\Tc$, in addition to the transition at $\Ts > \Tc$.
Fitting the $B_8$ susceptibility data near $\Tc$ to (\ref{chi ansatz}) with both $y_t$ and $y_h$ free, we find a second $B_8$ transition at $T=0.4316(4)$, with exponents $y_t=1.00(1)$ and $y_h=1.94(1)$.
The estimate $y_t=1.00(1)$ agrees perfectly with the exact Ising value $y_t=1$, and the transition temperature is also entirely consistent with our estimate of the chiral transition temperature $\Tc$.
We therefore conjecture that at the chiral transition temperature $\Tc$, the spins on the $B_8$ and $C_8$ sublattices undergo an Ising transition separating two QLR-ordered phases.
The value of $y_h=1.94(1)$ obtained from the fits of $\chi_{B_8}$ near $\Tc$ is clearly not equal to the usual Ising value of $y_h$ however; we therefore further conjecture that
this value is a newly observed critical exponent of the Ising universality class.
Determining the exact value of $y_h=1.94(1)$ remains to be further explored; we note however that from the Kac table, it may be $\frac{187}{96}\approx1.9479\ldots$.
For illustration, the inset of Fig.~\ref{UJ A4 and chiral} shows $\chi_{B_8} L^{-91/48}$ vs $T$.

\begin{figure}
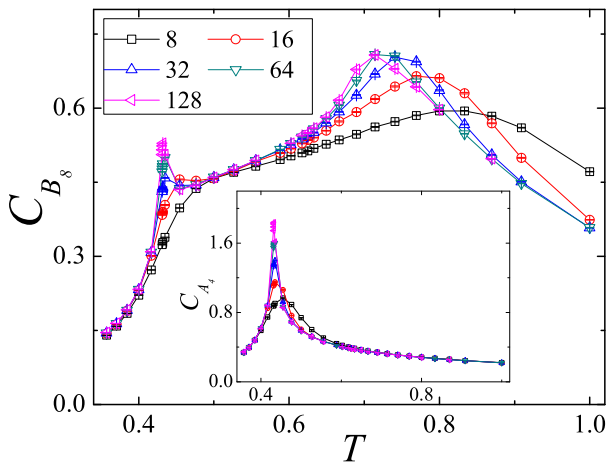

  \begin{center}
    \UJCBfig
    \caption{\label{UJ B8 specific heat} (Color online) $C_{B_8}$ versus $T$ on the UJ lattice. The inset shows $C_{A_4}$.}
  \end{center}
\end{figure}
\begin{figure}
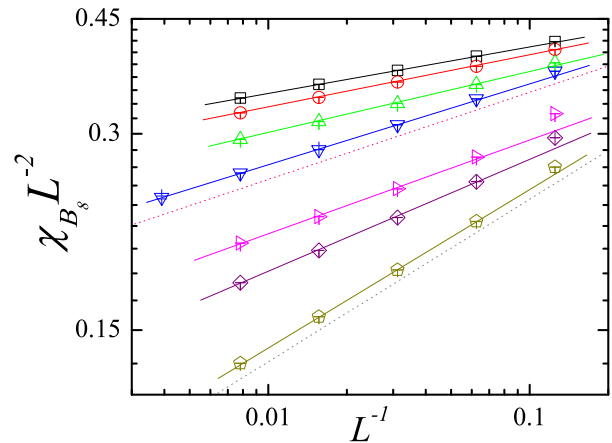

  \begin{center}
    \UJchiBfig
    \caption{\label{UJ B8 slopes} (Color online) Log-log plot of $\chi_{B_8}L^{-2}$ versus $L^{-1}$ on the UJ lattice. The dotted lines correspond to the critical points.}
  \end{center}
\end{figure}

As further evidence that the $B_8$ sublattice undergoes transitions at both $\Ts$ and $\Tc$, Fig.~\ref{UJ B8 specific heat} shows the specific-heat-like quantity $C_{B_8}$.
We clearly observe a peak near $\Ts$ and a divergence near $\Tc $ as $L \rightarrow \infty$. For comparison, the inset shows $C_{A_4}$, which does not show any peak near $\Ts$.
To illustrate that the phases of the $B_8$ spins on either side of $\Tc$ are indeed both QLR ordered,
we display in Fig.~\ref{UJ B8 slopes} a log-log plot of the $\chi_{B_8}L^{-2}$ data vs $L^{-1}$ for various values of $T$.
We clearly observe an algebraic divergence of $\chi_{B_8}$, with a $T$-dependent exponent.

\begin{figure}
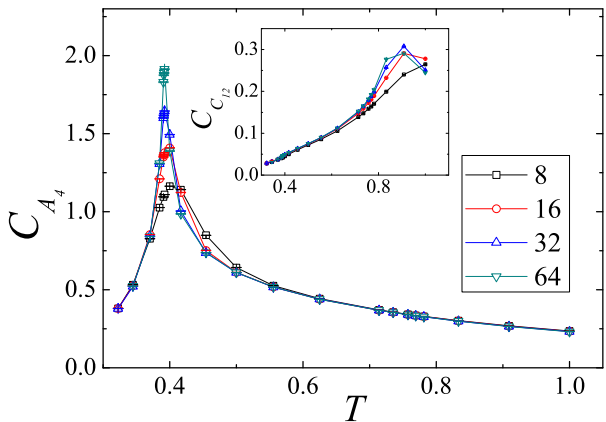

  \begin{center}
    \BHCfig
    \caption{\label{BH specific heat} (Color online) $C_{A_4} $ versus $T$ on the BH lattice. The inset is for $C_{C_{12}}$.}
  \end{center}
\end{figure}

\paragraph{Bisected-hexagonal lattice.---}
Analyzing $\chic$ and $\Qc$ in the same manner as for the triangular and UJ lattices, we find that the chiral order parameter undergoes an Ising transition at $\Tc=0.39137(8)$.
Similarly, an analysis of the sublattice susceptibilities shows that the spins on each sublattice undergo a transition from disorder to QLR order at a common value of $\Ts=0.747(2)$.
The magnetic transitions are again consistent with the KT universality class.
As for the $B_8$ and $C_8$ sublattices on the UJ lattice, $\chi_{A_4}$ and $\chi_{B_6}$ show that the $A_4$ and $B_6$ sublattices of the BH lattice also undergo an Ising transition at $T=\Tc$.
Near $\Tc$, the specific-heat-like quantities $C_{A_4}$ and $C_{B_6}$ are observed to be divergent for $L \rightarrow \infty$, while $C_{C_{12}}$ seems to be a smooth function.
By contrast, near $\Ts$, a peak occurs in $C_{C_{12}}$ which is absent in $C_{A_4}$ and $C_{B_6}$; see Fig.~\ref{BH specific heat}.
These observations suggest that the magnetic transition on $C_{12}$ is a generic KT transition,
and that the QLR magnetic order on $A_4$ and $B_6$ in the region $\Tc\le T\le\Ts$ is likely induced by the long-range correlation length on $C_{12}$.

\paragraph{Discussion.---}
In this Letter, we have studied the XY antiferromagnet on three plane triangulations: the triangular, UJ and BH lattices.
Each of these lattices is tripartite and Eulerian (all vertex degrees are even).
In each case, we find that the chiral order parameter undergoes a standard Ising order/disorder transition.
This behavior should be generic on all Eulerian tripartite triangulations of the plane, since the set of ground states will share the same $\mathbb{Z}_2$ chiral degeneracy.
However, as we have shown, the nature of the magnetic transitions appears to be strongly dependent on the specific lattice topology.

In addition to the intrinsic theoretical interest of these results, we note the AFXY model on the UJ lattice could be realized experimentally using Josephson Junction Arrays
in an entirely similar manner to the triangular-lattice model. Therefore, the results that we have outlined above should be experimentally observable.

\paragraph*{Acknowledgments.---} We are indebted to Nikolay Prokofiev, Boris Svistunov, Jon Machta and Henk Bl\"ote for valuable discussions.
This work was supported by NSFC (Grant Nos. 10975127 and 11147013), CAS, and the Fundamental Research Funds for the Central Universities (Grant Nos. 2012QNA43 and 2011RC25).
It was also supported under the Australian Research Council's Discovery Projects funding scheme (project number DP110101141),
and T.G. is the recipient of an Australian Research Council Future Fellowship (project number FT100100494).
T.G. is grateful for the hospitality shown by the University of Science and Technology of China,
and in particular the Hefei National Laboratory for Physical Sciences at Microscale, at which some of this work was completed.

\end{document}